\documentclass[11pt]{article}
\usepackage{moriond,epsfig}

\def\be{\begin{equation}}
\def\ee{\end{equation}}
\def\bea{\begin{eqnarray}}
\def\eea{\end{eqnarray}}

\begin{document}
\title{TIME-REVERSAL SYMMETRY BREAKING BY AC FIELD: EFFECT OF COMMENSURABILITY IN
THE
FREQUENCY DOMAIN
}

\author{\underline{V.E. KRAVTSOV}},

\address{The Abdus Salam International Centre for Theoretical 
Physics, P. O. Box 586, 34100 Trieste, Italy \\
and Landau Institute for Theoretical Physics, 2 Kosygina 
Street, 117940 Moscow, Russia}

\maketitle\abstracts
{
It is shown that the variance of the linear dc conductance fluctuations in
an open
quantum dot under a high-frequency ac pumping depends significantly
on the spectral content of the ac field. For a sufficiently strong ac
field the dc conductance
fluctuations are much stronger for the periodic pumping than in the case
of the noise ac field of the same intensity.
The reduction
factor $r$  in a  static magnetic field  takes the universal value of
2
only for the white--noise pumping. In general $r$ may deviate from 2 thus signalling
on
the time-reversal breaking by the ac field.
For the bi-harmonic 
ac field of the form
$A(t)=A_{0}\,[\cos(\omega_{1} t)+\cos(\omega_{2} t)]$ 
we predict the  enchancement 
of effects of $T$-symmetry breaking at commensurate frequencies
$\omega_{2}/\omega_{1}=P/Q$. In the high-temperature limit 
there is also the parity effect: the enchancement is only present if either $P$ or
$Q$ is even.
}

\section{Introduction}
Recently there has been a considerable interest in non-equilibrium
mesoscopics. The effect of adiabatic charge pumping \cite{Thou} has been
experimentally observed \cite{Marc} and analyzed theoretically
\cite{Brou,SAA}.
Weak localization under ac pumping \cite{VA}  and the
photovoltaic effect \cite{VAA} in a quantum dot have been
theoretically studied.
The non-equilibrium noise has been suggested \cite{krav3} as a cause of
both
the low
temperature dephasing saturation \cite{moh1} and the anomalously large
ensemble averaged persistent current \cite{Levi}.

Here we study the effect of the high-frequency ac field on the mesoscopic
fluctuations of {\it linear} dc conductance when both the {\it weak dc
voltage} and a {\it strong enough high-frequency pump field} are
applied to the
open quantum dot.
We will study the dependence of the variance of the dc conductance
fluctuations on the ac field intensity for  the noiselike and an almost
periodic ac field.
In particular we focus on the reduction factor   
$r=1+{\cal C}/{\cal D}$ for the variance of conductance
fluctuations after
turning on a strong  static magnetic  
field that kills the cooperon contribution to the variance $\langle
\delta G^2 \rangle_{C}\equiv{\cal C}$
while
leaving the diffuson one $\langle\delta G^2 \rangle_{D}\equiv {\cal D}$
unchanged.

This question is of the fundamental importance, since the universality of the
reduction factor is related with the time-reversal invariance of the system
without the static magnetic field. The external ac field certainly breaks  
the time-reversal invariance. However, if the characteristic frequency
$\omega$ of the ac field is high it is intuitively clear that the effect of
time-reversal symmetry breaking  is observed only at special conditions. 

Below we show that not only the power but also the spectral content of the 
ac field is important for an effective $T$-symmetry breaking. In particular, the
white ac noise is shown to be the most effective way to diminish the conductance 
fluctuations by dephasing. Yet it does not lead to any deviations from the   
universal value of $r=2$. The periodic ac field is least effective in dephasing
but it produces the maximum possible (at a given ac power) effect of time-reversal
breaking.

An interesting special case arises when the ac field is the superposition of two
harmonic fields with different frequencies $\omega_{1}$ and $\omega_{2}$. We will
show below that both the variance of conductance fluctuations and the
inverse reduction
factor $1/r$ have very sharp peaks as a function of the ratio
$\alpha=\omega_{2}/\omega_{1}$ at any rational value of $\alpha=P/Q$ but the height 
of the peak decreases with increasing the denominator $Q$. 
Thus we conclude that the
statistics of mesoscopic conductance fluctuations reveals {\it the effect of
commensurability in the frequency domain} on the dephasing and the time-reversal
breaking by the ac filed.
\section{The Landauer conductance in the time domain}
The
Landauer conductance $g=(\gamma V/4)^2 \,[K({\bf r},{\bf r'})+K({\bf
r'},{\bf r})]$ of a dot
of the volume $V$ with
small contacts at ${\bf r}$ and ${\bf
r'}$ and the electron escape rate $\gamma$, can be expressed in terms of
the exact retarded and advanced electron Green's
functions $G^{R,A}({\bf r},{\bf r'};t,t')$ in the time domain \cite{VA,YKK,WKr}:
\begin{eqnarray}
\label{cond}
K({\bf r},{\bf r'})=\int
dt_{1}dt_{2}\,\overline{G^{R}({\bf
r},{\bf
r'};t,t_{1})G^{A}({\bf
r'},{\bf r};t_{2},t)}\,F_{t_{1}-t_{2}}
\end{eqnarray}
where $F_{t}=\pi
Tt\sinh^{-1}(\pi T t)$ is the
Fourier-transform of
the derivative of the Fermi distribution function for electrons in the leads and
$\overline{f(t)}=\int_{-{\cal T}/2}^{{\cal T}/2}\frac{dt}{\cal T} f(t)$
denotes time averaging
during  the observation time ${\cal T}$.
\begin{figure}[tbp]
\centerline{\epsfysize=4.0cm \epsffile{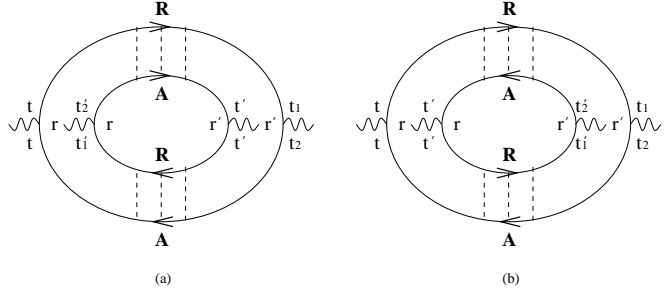}}
\caption{The cooperon (a) and the diffuson (b) contributions to the
variance
of conductance fluctuations.}
\end{figure}
We consider a chaotic or disordered quantum dot with the number of open
channels $M\gg 1$
and the electron
escape rate $\Delta \ll \gamma=\Delta M \ll E_{c}$
where $\Delta$ is the mean level separation
and the Thouless energy $E_{c}$ is the inverse ergodic time.
We also assume
the dephasing rate $\gamma_{int}(T)$ caused by electron interaction to be
smaller than
the escape rate $\gamma$.
In this situation the charging effects are negligible, the perturbative
diagrammatic analysis (see Fig.1) is possible and the ergodic zero-dimensional
approximation applies. 
In this approximation the diffuson and the cooperon contribution to the variance of
conductance fluctuations is given by \cite{WKr}:
\begin{equation}
\label{d2}
{\cal D}\propto\int dt^{\prime }dt^{\prime \prime }
\overline{D_{\eta}
(t,t^{\prime    
})D_{-\eta
}(t,t^{\prime \prime })}\;F^{2}_{t^{\prime }-t^{\prime \prime}}.
\end{equation}
\begin{eqnarray}
\label{c2}
{\cal C}\propto\int dt^{\prime}d\eta ^{\prime }
\overline{C_{t^{\prime}+t}
(-\eta^{\prime}-2t',-\eta+2t')C_{t}(\eta,\eta
^{\prime})}\,F^{2}_{2t^{\prime}}.
\end{eqnarray}
where averaging over $t$ and $\eta$ is assumed.

In Eqs.(\ref{d2}),(\ref{c2}) the time-dependent diffuson $D_{\eta}(t,t')$ and
cooperon $C_{t}(\eta,\eta')$ are certain products of electron retarded ($G^{R}$) or
advanced ($G^{A}$) Green's functions averaged over disorder:
\begin{equation}
\label{coop-dif}
\langle G^{R}(t_{+},t^{\prime}_{+})
G^{A}(t_{-}^{\prime},t_{-})
\rangle=\delta(\eta-\eta')\,D_{\eta}(t,t^{\prime }),\;\;
\langle G^{R}(t_{+},t_{+}^{\prime})
G^{A}(t_{-},t_{-}^{\prime})
\rangle =\frac{1}{2}\, \delta(t-t')\, C_{t}(\eta, \eta ^{\prime }),
\end{equation}
where $t_{\pm}=t\pm\eta/2,
t_{\pm}^{\prime}=t^{\prime}\pm\eta^{\prime}/2$.
\newpage
In the ergodic regime they are independent of the coordinates and are given by
\cite{YKK,WKr}:
\begin{equation}
\label{dif1}
D_{\eta}(t,t^{\prime
})=\Theta_{\eta-\eta'}\,\exp\left[-\int_{t'}^{t} \Gamma_{d}(\eta,\xi)\,d\xi\right]   
\end{equation}  
\begin{equation}
\label{coop1}
C_{t}(\eta,\eta^{\prime
})=\Theta_{t-t'}\,\exp\left[-\frac{1}{2}\int_{\eta'}^{\eta}
\Gamma_{c}(t,\xi)\,d\xi\right],
\end{equation}  
where $\Theta_{t}$ is the step-function.

The functions $\Gamma_{d}(\eta,\xi)$ and $\Gamma_{c}(t,\xi)$ describe the dephasing
caused by the ac field and by the electron escape into leads. For a  
ring with the diffusive electron motion and the 
circular electric field $E(t)=-\frac{\partial}{\partial t}A(t)$ cased by the
time-dependent flux  through it, these functions are given by \cite{WKr,YKK}:
\begin{equation}
\label{gamd}
\Gamma_{d}(\eta,\xi)=\gamma+D\left(A(\xi+\eta/2)-A(\xi-\eta/2)\right)^2,
\end{equation}
\begin{equation}
\label{gamc}
\Gamma_{c}(t,\xi)=\gamma+D\left(A(t+\xi/2)+A(t-\xi/2)
\right)^2.
\end{equation}
where $\gamma$ is an electron escape rate, $A(t)$ is the vector-potential, and $D$
is an electron diffusion coefficient.

The same expressions hold \cite{YKK} for a single connected dot 
with a homogeneous longitudinal electric field $E(t)$ (the vector-potential
$A(t)$ is unambiguously defined by the condition
$\overline{A(t)}=\overline{E(t)}=0$) but only in
the high-frequency limit $\omega\gg E_{c}$. In the adiabatic limit $\omega\ll E_{c}$
one obtains \cite{VA,YKK}
\begin{equation}
\label{ad-d}
\Gamma_{d}(\eta,\xi)=\gamma+C
(E(\xi+\eta/2)-E(\xi-\eta/2))^2 , 
\end{equation}
\begin{equation}
\label{ad-c}
\Gamma_{c}(t,\xi)=\gamma+(C/4)
(E(t+\xi/2)-E(t-\xi/2))^2 
\end{equation}
where $C\sim L^{2}/E_{c}$, $L$ being the dot size. 

The existence of two different forms Eqs.(\ref{gamd}),(\ref{gamc}) and
Eqs.(\ref{ad-d}),(\ref{ad-c}) is the manifestation of the fact \cite{YKK} that
there are  two different {\it time-dependent} random matrix theories
for each Dyson universality class.
\section{The limit of high frequencies}
Eqs.(\ref{d2}),({\ref{c2}}) can be significantly simplified in the limit of high
frequencies $\omega\gg\gamma$.  At such a high-frequency pumping one can replace
$\Gamma_{d}(\eta,\xi)$ and $\Gamma_{c}(t,\xi)$ in
Eqs.(\ref{dif1},\ref{coop1}) by the
time averages
$\Gamma_{d}(\eta)=\overline{\Gamma_{d}(\eta,\xi)}$ and
$\Gamma_{c}(t)=\overline{\Gamma_{c}(t,\xi)}$. Then we obtain:
\begin{eqnarray}
\label{LowTd}   
\frac{{\cal
D}}{g_{0}^2}=\int_{-{\cal
T}/2}^{{\cal T}/2}
\frac{d\eta}{2{\cal
T}}\,\left(\frac{\gamma^2}{\Gamma_{d}(\eta)} \right)
\int_{0}^{\infty}e^{-t\,\Gamma_{d}(\eta)}\,F^{2}_{t}\,dt.
\end{eqnarray}
\begin{eqnarray}
\label{LowTc}   
\frac{{\cal C}}{g_{0}^2}=
\int_{-{\cal
T}/2}^{{\cal T}/2}\frac{dt}{{\cal T}}\int_{-\infty}^{t}dt'\,
\frac{\gamma^2 F^{2}_{2t-2t'}\,e^{-2(t-t')\,\Gamma_{c}(t')}}
{\frac{1}{2}(\Gamma_{c}(t)+\Gamma_{c}(t'))},
\end{eqnarray}
where $g_{0}=\pi\gamma/4\Delta$ is the mean conductance, 
${\cal T}$ is the observation time, and and $F_{t}=\pi T
t \sinh^{-1}(\pi T t)$ is the Fourier-transform of the Fermi distribution function.

In Fig.2 we show the result of a direct numerical evaluation of integrals in 
Eqs.(\ref{d2}),({\ref{c2}}) at $T=0$. One can see how the high-frequency limit is
reached  for the inverse reduction factor in the case of harmonic pumping. 
\begin{figure}[tbp]
\centerline{\epsfysize=5.0cm \epsffile{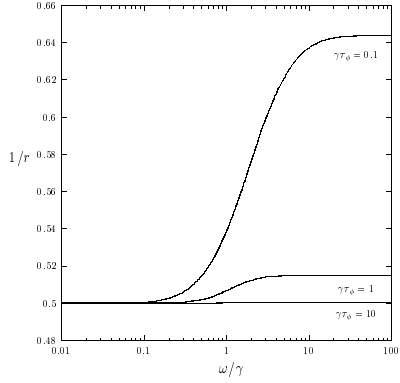}}
\caption{The frequency dependence of the inverse reduction factor for the harmonic
pumping at different dimensionless intensities $I= 1/\gamma\tau_{\varphi}$ and low 
temperature $T\ll \gamma$.}
\end{figure}
Eqs.(\ref{LowTd},\ref{LowTc}) can be further simplified in the limit of low
temperatures $T\ll\gamma$ where $F_{t}\approx 1$ and in the limit of high
temperatures $T\gg\omega(\gamma\tau_{\varphi})^{-1/2}$ where
$F^2_{t}\approx (\pi/6T)\,
\delta(t)$ and we have:
\begin{equation}
\label{highT}
\frac{\langle \delta G^2\rangle_{D,C}}{g_{0}^2}=\frac{\pi \gamma^2}{12 T}
\int_{-{\cal T}/2}^{{\cal T}/2}\frac{dt}{{\cal
T}}\,\frac{1}{\Gamma_{d,c}(t)}.
\end{equation}
Below we consider only these limiting cases. 

\section{Conductance fluctuations for the noise and harmonic ac pumping} 
There is a dramatic difference between the noiselike ac
field with
the short correlation time $\tau_{0}\sim \omega^{-1}\ll \tau_{\varphi}$
and
the harmonic ac field $A(t)=A_{0}\cos(\omega t)$ with $\omega\gg   
\gamma$. 
In the case of the white--noise pumping the time-average of the cross-terms
$\overline{A_{\xi+\eta/2}A_{\xi-\eta/2}}$ and
$\overline{A_{t+\xi/2}A_{t-\xi/2}}$
in Eqs.(\ref{gamd},\ref{gamc}) or Eqs.(\ref{ad-d}),(\ref{ad-c}) is zero and we
obtain {\it the same,
time-independent} dephasing
rates
$\Gamma_{d}=\Gamma_{c}=\gamma+1/2\tau_{\varphi}$
for the cooperons and the diffusons with $\eta\neq 0$,
where
\begin{equation}
\label{phi}
\tau_{\varphi}^{-1}=\left\{ \matrix{4 D \overline{A^2}& \omega\gg E_{c}\cr
4C \overline{E^2} & \omega\ll E_{c}\cr}\right.
\end{equation}
is proportional to the pumping intensity. For convenience of the further analysis we 
introduce the {\it dimensionless intensity} $I$:
\begin{equation}
\label{I}
I= \frac{1}{\tau_{\varphi}\gamma}
\end{equation}
which for the harmonic pumping is equal to the number of absorbed/emitted field
quanta $\hbar\omega$ for the time $\gamma^{-1}$ electron spends inside the dot.

Then Eqs.(\ref{d2}),(\ref{c2}) give ${\cal C}={\cal D}$, that is the diffuson and
the cooperon contributions to the variance of conductance fluctuations are equal to
each other. Hence {\it the time-reversal symmetry is effectively unbroken}.

The variance of conductance fluctuations in this case is given by:
\begin{equation}
\label{noi}
\frac{{\cal D}}{g_{0}^2}=\frac{{\cal
C}}{g_{0}^2}=\left\{\matrix{\frac{1}{2}\,
\left(1+\frac{I}{2}
\right)^{-2}, & T\ll \gamma +\frac{1}{2\tau_{\varphi}}\cr
\frac{\pi\gamma}{12 T}\,\left(1+\frac{I}{2}  
\right)^{-1}, &  T\gg \gamma +\frac{1}{2\tau_{\varphi}}\cr
}\right.   
\end{equation} 
For the  harmonic pumping the dephasing rates are periodic functions of time:
\begin{equation}
\label{deffun}
\Gamma_{d}(\eta)\equiv
\overline{\Gamma_{d}(\eta,\xi)}=\gamma+\tau_{\varphi}^{-1}\,\sin^2
(\omega \eta/2),\;\;\; 
\Gamma_{c}(t)\equiv \overline{\Gamma_{c}(t,\xi)}=\gamma+\tau_{\varphi}^{-1}\,\cos^2
(\omega t), 
\end{equation}
and
Eqs.(\ref{LowTd},\ref{LowTc}) no longer lead to the same result.

For large pumping intensities $I=(\tau_{\varphi}\gamma)^{-1}\gg 1$
the dephasing functions Eq.(\ref{deffun}) are small in the {\it `no-dephasing'
windows} near zeros $\eta_{n}$ and $t_{n}$ of $\sin^2 (\omega \eta/2)$ and
$\cos^2(\omega
t)$.
It is these no-dephasing windows of the width $\omega^{-1} I^{-1/2}$ that make the
dominant contribution to the
magnitude of all phase-coherent
effects at high pumping intensities $I\gg 1$.

In order to compute the diffuson and the cooperon contribution to the conductance
fluctuations one can expand $\Gamma_{d}(\eta)$ and $\Gamma_{c}(t)$ in
Eqs.(\ref{LowTd}),(\ref{LowTc}) near
$\eta_{n}$, and $t_{n}$,   
perform integrations from $-\infty$ to $+\infty$ over $\delta
\eta=\eta-\eta_{n}$ and $\delta t = t-t_{n}$
and sum over all $n$. The result depends on the relation between
the temperature of leads $T$ and $\gamma$.
For $T\ll \gamma\ll \tau_{\varphi}^{-1}$ we obtain:
\begin{equation}
\label{dd}
\frac{{\cal D}}{g_{0}^2}\approx
\frac{1}{4\sqrt{I}},\;\;\;\;\;\;\;\;\;\;\;\frac{{\cal
C}}{g_{0}^2}\approx \frac{1}{2I},\;\;\;\;\;\;\;\;\;\;\;(I\gg 1).
\end{equation}
One can see that the diffuson contribution ${\cal D}$ to the conductance fluctuations 
is parametrically larger than the cooperon one ${\cal C}$ even in the absence of
a static magnetic field. This signals on the extremely strong $T$-breaking effect
in this regime with the reduction factor $r=1+2 I^{-1/2}$ close to 1.

However, it is enough to raise the temperature of leads to make the $T$-breaking
effect negligible. 
At high temperatures $T\gg \omega\sqrt{I}$ and strong
pumping $I\gg 1$ Eq.(\ref{highT}) gives the same
value
\begin{equation}
\label{HT}
\frac{{\cal D}}{g_{0}^2}=\frac{{\cal C}}{g_{0}^2}=\frac{\pi\gamma}{12
T}\;\frac{1}{\sqrt{I}}.
\end{equation}
for the diffuson and the cooperon contribution to conductance
fluctuations, and the reduction factor $r=2$.

Comparing Eq.(\ref{noi}) with Eqs.(\ref{dd}),(\ref{HT}) we also conclude
that at high intensities $I\gg 1$ the suppression of mesoscopic fluctuations
by the noiselike ac field is always stronger than the suppression by a harmonic
field of the same power. This is because the `no dephasing' windows are absent
in the case of noise.
\section{Conductance fluctuations for an almost periodic ac field and statistics of
zeros of the dephasing functions.}
The idea of the `no-dephasing windows' in the vicinity of zeros
of the dephasing functions $\Gamma_{d}(\eta)-\gamma$ and $\Gamma_{c}(t)-\gamma$ can
be applied to a
general case of an almost periodic ac field. For such an ac filed the dephasing
functions have a certain density of {\it complex} zeros $z_{n}=x_{n}+iy_{n}$ with a
small
complex part $y_{n}$:
\begin{eqnarray}
\label{df}
\rho(y)=  
\sum_{n}\langle\delta(x-x_{n})\delta(y-y_{n})\rangle_{x},
\end{eqnarray}
where $\langle
...\rangle_{x}=({\cal T})^{-1}\int_{|x|<{\cal
T}/2}dx ...$  stands for the averaging
over $x$.

Since only the time variables in the `no dephasing windows' contribute to
the variance of mesoscopic fluctuations at large pumping intensities, one can 
express the dependence of the variance on the dimensionless intensity
$I$ in terms of the density of zeros $\rho(y)$.

For an important example of the {\it bi-harmonic} ac field
\begin{equation}
\label{bi-harm}
A(t)=A_{0}\,[\cos(\omega t)+ \cos(\alpha \omega t)]
\end{equation} 
one obtains in the high-temperature limit \cite{WKr} $T\gg \omega \sqrt{I}$:
\begin{equation}
\label{DD}
\frac{\langle \delta G^2
\rangle_{D,C}}{g_{0}^2}=\frac{\pi^2\gamma 
}{3T \sqrt{2(1+\alpha^2)}}\,\,I^{-1}\int_{-\infty}^{+\infty}\frac{  
dy\, \rho_{d,c}(y)}{\left[I^{-1}+
\frac{y^2}{8}\,(1+\alpha^2)
\right]^{1/2}},
\end{equation}
where $\rho_{d,c}$ is the density of zeros that corresponds to the 
equation (the sign $\pm$ stands for the cooperon(diffuson) part):
\begin{equation}
\label{eq}
\cos z + \cos(\alpha z)\pm 2 =0.
\end{equation} 
\section{Effect of commensurability in the frequency domain}
There is a drastic difference in the density of zeros $\rho_{d,c}(y)$ for the case
of {\it commensurate} and {\it incommensurate} frequencies $\omega_{1}=\omega$ and
$\omega_{2}=\alpha\omega$ in Eq.(\ref{bi-harm}). For the case of 
commensurate
frequencies
$\alpha=P/Q <1$ the density $\rho(y)$ is the set of
$\delta$-functions $(2\pi Q)^{-1}\sum \delta (y-y_{n})$ separated by
gaps
$\Delta y_{n}\sim 1/Q$. In the limit of incommensurate frequencies
$Q\rightarrow\infty$ the function $\rho(y)$ is continuous with no singularity at
$y=0$. This difference shows up in the dependence of the diffuson
and the
cooperon
contribution to the variance of conductance fluctuations on the pumping intensity
$I$. 

Let us consider the case of commensurate frequencies $\alpha=P/Q$ with not very
large  denominator $Q$. 
At large dimensionless intensities $I\gg Q^2$ the gap
$y_{1}\sim 1/Q$ is large compared to $I^{-1/2}$, and
one can neglect in Eq.(\ref{DD}) all complex roots with
$y_{n}\neq 0$. Equation (\ref{eq}) with the sign minus has real roots at {\it any}
$\alpha$. Then one immediately concludes from Eq.(\ref{DD}) that for  commensurate
frequencies with $Q \ll \sqrt{I}$ the diffuson part of the variance of conductance
fluctuations is proportional to $I^{-1/2}$. However,
Eq.(\ref{eq}) with the sign plus
(relevant for the cooperon contribution) has real solutions
only if $P$ and $Q$ are {\it both odd}.
In this case in the high-temperature limit we have ${\cal D}={\cal C}\propto
I^{-1/2}$ as
for a strictly harmonic pumping.  
Thus the {\it symmetry between the diffuson and the cooperon contribution is
unbroken}.
One can see that the condition that both $P$ and $Q$ in
$\alpha=P/Q$ in Eq.(\ref{bi-harm}) are odd,
is a particular case of  a
more general condition \cite{Flach}
\begin{equation}
\label{antisym}
A(t+\tau^{*})=-A(-t+\tau^{*}), 
\end{equation}
that is, by a certain shift  $\tau^{*}$ (in our case $\tau^{*}$ is a quarter of 
the period)  the vector-potenial $A(t)$ can be made
an odd function of time. This is the natural definition  for the
time-reversal symmetry in the absence of a preferential time origin. 

If
either $Q$ or $P$ is even this condition is violated and
the cooperon contribution is
anomalously suppressed  ${\cal C}\propto I^{-1}$. In this case the effective
$T$-symmetry breaking is strong
even in the high-temperature limit (see Fig.4. and Fig.5.). 
Such a  {\it parity effect} is
also present in the
low-temperature limit $T\ll
\gamma$ as it is seen from
Fig.3 obtained by numerical
integration of Eqs.(\ref{LowTd},\ref{LowTc}). However, in this case it is much
weaker and the symmetry Eq.(\ref{antisym}) no longer implies ${\cal C}={\cal D}$. 
\begin{figure}[tbp]
\centerline{\epsfysize=6.0cm \epsffile{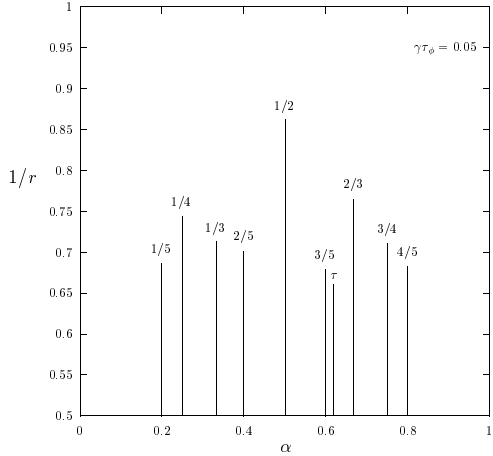}}
\caption{ Reduction factor $r$ vs the frequency ratio $\alpha$ 
in simple commensurate points $\alpha=P/Q$ for low temperatures $T\ll \gamma$ and 
pumping intensity $I=(\gamma\tau_{\varphi})^{-1}=20$. Also plotted is $r$ for the
golden mean
$\alpha=\tau=\frac{\sqrt{5} -1}{2}$.}
\end{figure}

Now consider the case $Q\gg \sqrt{I}$. In this case the function $\rho_{d,c}(y)$ 
in Eq.(\ref{DD})is effectively averaged over the interval $\delta y \sim I^{1/2}$
and can be replaced by a constant.
As the result both the diffuson and the cooperon part of the variance is
proportional to $I^{-1}\,\ln I$ and the effect of $T$-breaking drastically
decreases. That is why the depth of dips in the reduction factor $r$ at $\alpha=P/Q$
rapidly decrease with increasing the denominator $Q$
at a given intensity $I$ (see Fig.4 and Fig.5.). The same trend holds for the
low-temperature regime too (Fig.3) but the reduction factor $r$ does not reach its
universal value even for the `most irrational' frequency ratio
$\alpha=(\sqrt{5}-1)/2$. 

\begin{figure}[tbp]
\centerline{\epsfysize=5.4cm \epsffile{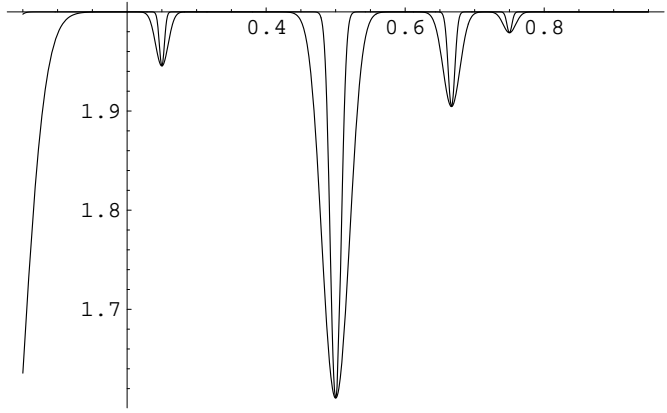}}
\caption{The reduction factor $r$ as a function of
the frequency ratio $\alpha$ in the high-temperature limit $T\gg \omega\sqrt{I}$ for
$I=4$ and
$1/{\cal T}\omega=0.05$ and 0.02. The dips at $\alpha=\frac{1}{2},
\frac{2}{3},\frac{1}{4}$
and
$\frac{3}{4}$ are seen while there is no dip for $\alpha=\frac{1}{3}$ where
both $P$ and $Q$ are odd. The width of peaks is of the order of $1/{\cal
T}\omega$.}
\end{figure}
\begin{figure}[tbp]
\centerline{\epsfysize=6.4cm \epsffile{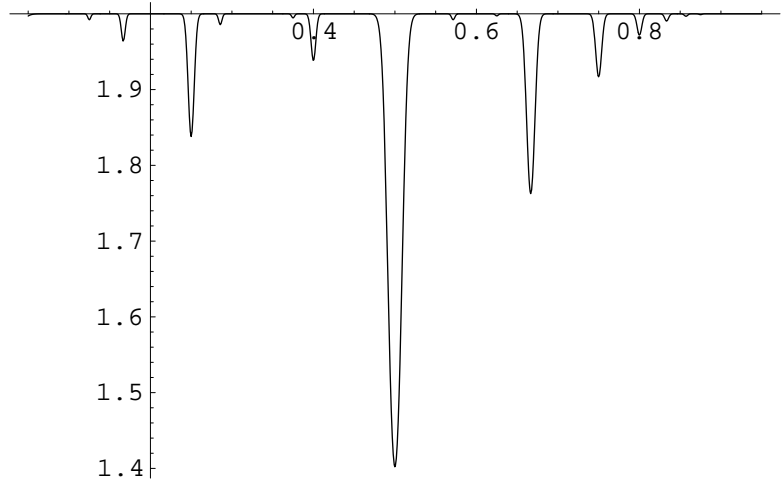}}
\caption{The reduction factor $r$ as a function of
the frequency ratio $\alpha$ in the high-temperature limit $T\gg \omega\sqrt{I}$ for
$I=8$ and
$1/{\cal T}\omega=0.02$. Small dips at $\alpha=P/Q$ with larger denominator $Q$ 
($PQ$ is even)
become visible.}
\end{figure}
\begin{figure}[tbp]
\centerline{\epsfysize=6.4cm \epsffile{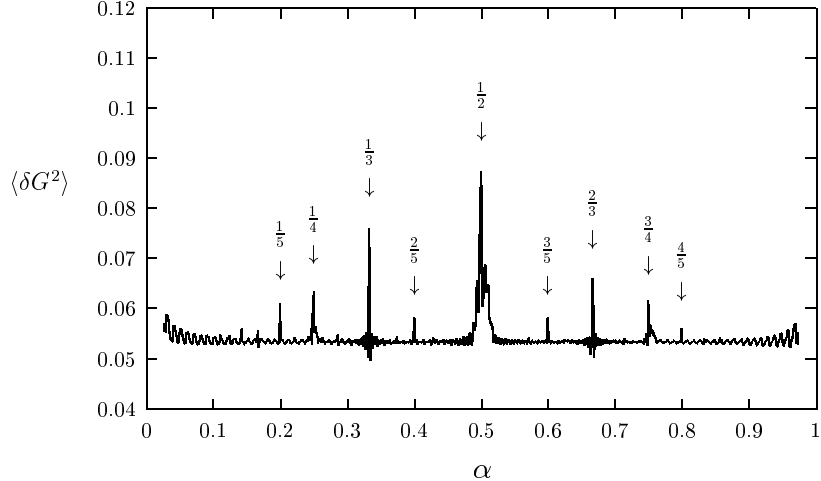}}
\caption{The total variance of conductance fluctuations (in units of the  
ensemble-average dc conductance) at $T\ll \gamma$ as a function of the frequency
ratio
$\alpha$ for $I=20$ and
$1/{\cal T}\omega=0.01$. The width of peaks is of the order of $1/{\cal T}\omega$.}
\end{figure}

It is 
possible to obtain an explicit analytical expression for the cooperon and the
diffuson contribution to the variance of conductance fluctuations in the
high-temperature limit $T\gg \omega\sqrt{I}$: 
\begin{equation}
\label{analD}
\frac{{\cal D}}{g_{0}^2}\approx \frac{\gamma}{6T}\sum_{n,m=-\infty}^{+\infty}\,
\tilde{\delta}(n-\alpha m)\,\left\{\matrix{I^{-1} K_{0}\left(2\sqrt{\frac{n^2
+m^2}{I}}\right), & n^2+ m^2 >0 \cr 
I^{-1}\ln(2\sqrt{I}), & m=n=0 \cr} \right. ,
\end{equation}
\begin{equation}
\label{analC}
\frac{{\cal C}}{g_{0}^2}\approx \frac{\gamma}{6T}\sum_{n,m=-\infty}^{+\infty}\, 
(-1)^{n+m}\,\tilde{\delta}(n-\alpha m)\,\left\{\matrix{I^{-1} 
K_{0}\left(2\sqrt{\frac{n^2
+m^2}{I}}\right), & n^2+ m^2 >0 \cr
I^{-1}\ln(2\sqrt{I}), & m=n=0 \cr} \right.,
\end{equation}
where $K_{0}(x)$ is the Bessel function and $\tilde{\delta}(\omega x)$ is the
spectral lineshape function of the harmonic component normalized so that
$\tilde{\delta}(0)=1$. Fig.4 and Fig.5 are obtained from
Eqs.(\ref{analD}),(\ref{analC}) for the Gaussian lineshape
$\tilde{\delta}(x)=e^{-x^2 \omega^2 {\cal T}^2}$. 
Eqs.(\ref{analD}),(\ref{analC}) reveal an important and a bit counter-intuitive
feature of the $\alpha$ dependences. The electron escape rate $\gamma$ enters only
in the dimensionless intensity $I$ (see Eq.(\ref{I})) that controls the magnitude
of peaks/dips in the $\alpha$-dependences. The width of peaks/dips is determined
by the spectral linewidth $\delta={\cal T}^{-1}$ of the harmonic components and can
be much smaller than $\gamma/\omega$. 

The commensurability effect is present not only for the reduction factor $r$
but also for the total variance of conductance fluctuations.
In Fig.6. the total variance is plotted as a function of the frequency ratio
$\alpha$ for the case of low temperatures $T\ll \gamma$.

In conclusion, the effect of ac pumping on the statistics of conductance
fluctuations possesses an interesting dualism: the periodic ac fields that are least
effective in suppressing of conductance fluctuations 
by dephasing, turn out to be the most effective in breaking the time-reversal
symmetry. An important exception is the periodic ac field $A(t)$ that 
obeys the symmetry Eq.(\ref{antisym}). At high temperatures its
effect on
the reduction factor $r$ is negligible while the conductance fluctuations are much
larger than for the white--noise of the same power.      

I am grateful to X.-B. Wang, V.I.Yudson, E.Kanzieper and V.M.Akulin and
O.M.Yevtushenko for a fruitful
collaboration on the problem.

\end{document}